\documentstyle[12pt]{article}
\begin{document}

\title{THE INERTIA OF HEAT AND ITS ROLE IN THE DYNAMICS OF DISSIPATIVE COLLAPSE}
\author{L. Herrera\thanks{Postal address: Apartado 80793, Caracas 1080 A, Venezuela; e-mail address:
laherrera@cantv.net.ve}
\\
Escuela de F\'\i sica, Facultad de Ciencias,\\
Universidad Central de Venezuela,\\
Caracas, Venezuela.\\
}
\date{}
\maketitle

\begin{abstract}
The  decreasing of the inertial mass density, established in the past for dissipative fluids,
is found to be produced by the ``inertial'' term of the transport equation. Once the transport equation is coupled to the dynamical equation one finds that  the contribution of the
inertial term  diminishes the effective inertial mass and  the ``gravitational'' force term, by the same factor. An intuitive picture, and prospective applications of this result to 
astrophysical scenarios are discussed. 
\end{abstract}
\newpage
\section{Introduction}
In 1930 Tolman \cite{Tolman} published a paper with the suggestive title ``On the weight of heat and thermal equilibrium in general relativity''. The very idea underlying Tolman's work
is very simple indeed, namely: since according to special relativity all forms of energy have inertia, this should also apply to heat. Therefore, because of the equivalence principle,
there should be also some weight associated to heat, and one should expect that thermal energy tends to displace to regions of lower gravitational potential. This in turn implies that
the condition of thermal equilibrium in the presence of a gravitational field must change with respect to its form in  absence of gravity. Thus a temperature gradient is necessary
in thermal equilibrium in order to prevent the flow of heat from regions of higher to lower gravitational potential. This result was confirmed some years later by Eckart and Landau and
Lifshitsz \cite{Eckart}. Indeed, in the transport equation derived by these authors, the ``inertial'' term deduced by Tolman appears explicitly.

However, we know that for systems out of equilibrium  or quasi--equilibrium (thermal and dynamic), the Landau-Eckart approach leads to difficulties \cite{6}, which can be solved only in
the context of a hyperbolic theory of dissipation \cite{Muller67}. Accordingly in order to bring out  the role of the inertia of heat in a dynamical
situation we have to consider a transport equation derived from the Israel--Stewart theory.

In a recent paper \cite{HS} we have derived the influence of dissipative processes in the collapse of a self--gravitating sphere. As a main result of that study it appears  that
the effective inertial mass density of a  fluid element  reduces by a factor which depends on dissipative variables. This result was already known (see \cite{eim} and references
therein), but considered to be valid  only, just after leaving the equilibrium, on a time scale of the order of relaxation time. The novelty in \cite{HS} is, on the one hand, that such
reduction of the effective inertial mass density is shown to be valid all along the evolution, and on the other, that the``gravitational force'' term in the dynamical equation
is also  reduced by the same factor, as expected from the equivalence principle (this is absent in  \cite{eim}
due to the fact that in that reference, equations are evaluated in a locally Minkowski frame, where local gravitational effects vanish).

However, the very origin of this effect has remained obscure until now.
Here  we shall show that such  decreasing of the effective inertial mass term is directly related to the inertial term in the transport equation. We shall further provide an intuitive
picture on how such decreasing comes about.
\section{The dynamical equation of the dissipative fluid}
We consider a spherically symmetric distribution of collapsing
fluid, which, for sake of completeness, we assume to be locally anisotropic,
undergoing dissipation in the form of heat flow, bounded by a
spherical surface $\Sigma$. 
We assume the interior metric to $\Sigma$ to be comoving, shear free for simplicity, and spherically symmetric, accordingly it may be written as
\begin{equation}
ds^2=-A^2(t,r)dt^2+B^2(t,r)(dr^2+r^2d\theta^2+r^2\sin^2\theta d\phi^2), \label{3}
\end{equation}
and hence we have for the four velocity $V^{\alpha}$ and the heat flux vector $q^{\alpha}$
\begin{equation}
V^{\alpha}=A^{-1}\delta^{\alpha}_0, \;\; q^{\alpha}=q\delta^{\alpha}_1, \;\; 
, \label{4}
\end{equation}

Then it can be shown \cite{HS}  that the following equation can be found
from Bianchi identities (since we are considering here  dissipation only  at diffusion approximation, we put $\epsilon =0$ )
\begin{eqnarray}
(\mu+P_r)D_tU=-(\mu+P_r)\left[m+4\pi P_r R^3 \right]\frac{1}{R^2} \nonumber \\
-E^2
\left[D_R P_r+2\frac{P_r-P_{\perp}}{R}\right] \nonumber \\
-E\left[5qB\frac{U}{R}+BD_tq\right].
\label{29}
\end{eqnarray}  
Where $\mu$ is the energy density, $P_r$ the radial pressure,  $P_{\perp}$ is the tangential
pressure, 
\begin{equation}
D_t=\frac{1}{A}\frac{\partial}{\partial t}. \label{16} 
\end{equation}
and  the velocity $U$ of the collapsing fluid is defined as
\begin{equation}
U=rD_tB<0 \;\; (in \; the \;  case \; of \; collapse). \label{19}
\end{equation}
Also, the mass function $m(t,r)$ of Cahill and McVittie \cite{Cahill} is obtained from the Riemann tensor component ${R_{23}}^{23}$ and is for metric (\ref{3})
\begin{equation}
m(t,r)=\frac{(rB)^3}{2}{R_{23}}^{23}=\frac{r^3}{2}\frac{B\dot{B}^2}{A^2}-\frac{r^3}{2}
\frac{B^{\prime 2}}{B}-r^2B^{\prime}, \label{8a}
\end{equation}
$E$ is defined as 
\begin{equation}
E=\frac{(rB)^{\prime}}{B}=\left[1+U^2-\frac{2m(t,r)}{rB}\right]^{1/2}, \label{20}
\end{equation}
and the proper radial derivative $D_R$,
constructed from the radius of a spherical surface, as measured from its perimeter inside $\Sigma$, being
\begin{equation}
D_R=\frac{1}{R^{\prime}}\frac{\partial}{\partial r}. \label{23a}
\end{equation}
 where 
\begin{equation}
R=rB, \label{23aa}
\end{equation}
and where dots and primes denote derivatives with respect to $t$ and $r$ respectively.

Next, the corresponding  transport equation for the heat flux reads \cite
{Muller67} 
\begin{equation}
\tau
h^{\alpha\beta}V^{\gamma}q_{\beta;\gamma}+q^{\alpha}=-\kappa h^{\alpha\beta}
(T_{,\beta}+Ta_{\beta}) -\frac 12\kappa T^2\left( \frac{\tau
V^\beta }{\kappa T^2}\right) _{;\beta }q^\alpha ,  \label{21}
\end{equation}
where $h^{\mu \nu }$ is the projector onto the three space orthogonal to $%
V^\mu $,  
$\kappa $  denotes the thermal conductivity, and  $T$ and  $\tau$
denote temperature and relaxation time
respectively. 
Observe that due to the symmetry of the problem, equation (\ref{21}) only has one independent component, which may be written as:

\begin{eqnarray}
BD_tq=-\frac{\kappa T}{\tau E}D_tU-\frac{\kappa T^{\prime}}{\tau B}-  
\frac{qB}{\tau}(1+\frac{\tau U}{R})-\nonumber \\
-\frac{\kappa
T}{\tau E}\left[m+4\pi P_r R^3\right]R^{-2}- 
\frac{\kappa T^2 q B}{2A\tau}\left(\frac{\tau}{\kappa T^2}\right)\dot{} -\frac{3U Bq}{2R}, \label{V3}
\end{eqnarray}
where the equation 
\begin{equation}
D_tU=-\left[m+4\pi P_r  R^3\right](R)^{-2}+\frac{A^{\prime}}{A}
\frac{(rB)^{\prime}}{B^2}, \label{27}
\end{equation}
has been used.

Then coupling (\ref{29}) to (\ref{V3}) one obtains (some misprints in
eq.(39) in \cite{HS} has been corrected here)
 \begin{eqnarray}
(\mu+P_r)(1-\alpha)D_tU=F_{grav}(1-\alpha)+F_{hyd}+\nonumber \\
+\frac{E\kappa T^{\prime}}{\tau B}+\frac{EqB}{\tau}-\frac{5qBE U}{2R} 
+\frac{\kappa ET^2 q B}{2A \tau}\left(\frac{\tau}{\kappa T^2}\right)\dot{}. \label{V4}
\end{eqnarray}
Where $F_{grav}$ and $F_{hyd}$ are defined by 
\begin{equation}
F_{grav}=-(\mu+P_r)\left[m+4\pi P_r R^3\right]\frac{1}{R^2}, \nonumber \\
\label{grav}
\end{equation}
and
\begin{equation}
F_{hyd}=
-E^2
\left[D_R P_r+2\frac{P_r-P_{\perp}}{R}\right],
\label{hyd}
\end{equation}
where $\alpha$ is given by
\begin{equation}
\alpha=\frac{\kappa T}{\tau (\mu+P_r)}.
\label{alpha}
\end{equation}

Now, interpreting (\ref{V4}) as  a  ``Newtonian'' equation of the form 
\begin{equation}
Force= Mass \; density \times Acceleration
\label{Newton}
\end{equation}
it appears that 
\begin{equation}
(\mu+P_r)(1-\alpha)
\label{EIM}
\end{equation}
plays the role of the effective inertial mass density.
Thus as $\alpha$ tends to $1$, the effective inertial mass density of the fluid element tends to zero.
Furthermore observe that $F_{grav}$ is also multiplied by the factor $(1-\alpha)$. Indicating that the effective
gravitational attraction on any fluid element decreases by the same factor as the effective inertial mass (density). Which of
course is to be expected, from the equivalence principle. It is also worth mentioning that $F_{hyd}$ is in principle independent
(at least explicitly) on this factor.

We shall next find the origin of such decreasing in the inertial mass term.
\section{Inertia of heat in the dynamic regime}
Let us now track the origin of the $\alpha$ terms  in (\ref{V4}). With this purpose in mind observe that the $Ta_{\beta}$ term on the right of (\ref{21}) represents the Tolman
inertial term. This in turn yields, using (\ref{27}), the first and fourth terms on the right of (\ref{V3}). The sign of the former depends on the sign of $D_tU$, whereas the
sign of the latter is always negative, reflecting the fact that thermal energy, due to its inertia, will tend to displace to the center, decreasing thereby any outwardly directed heat
flow.  But these are the terms producing the
$\alpha$ terms in (\ref{V4}). Thus any decreasing in the inertial mass term  in the dynamic equation, produced by an increasing of $\alpha $, is due to the
inertia of heat.

However, it does not appear physically clear yet, why the inertia of heat induces a decreasing of the inertial mass
term, in the dynamic regime. In order to understand this point, observe that as an element of fluid contracts  ($D_tU<0$), the mere existence of the inertia of heat will induce a
$D_tq>0$ (directed outward) as shown in (\ref{V3}). Furthermore, it appears that the inertial term in the expression for $D_tq$
produces a term proportional to $D_tU$ and a term proportional to the gravitational  force term.
From (\ref{29}) we see that a positive $D_tq$ produces an inwardly directed force, which as we have just said is proportional to $D_{t}U$. Thus, passing this term to the left hand side
of (\ref{V4}), such a term implies a decreasing of the total inertial mass.

At a more intuitive level, one might say that  the inertia of heat produces a ``laging  behind'' of the thermal energy with respect to
the contracting fluid element which  induces a
$D_tq>0$, and thereby a decreasing in the inertial mass of the fluid element. The decreasing in the gravitational force term is a direct consequence of the equivalence principle.

\section{Conclusions}
We have seen that the inertia of heat, which atracted the attention of Tolman many years ago in relation with the thermal equilibrium condition, may play an important role in the
dynamics of a self--gravitating object, by decreasing the inertial mass  and the gravitational force term.

 Indeed, as far as the right hand side of (\ref{V4}) is negative, the system
keeps collapsing. However, if the  collapsing sphere evolves in such a way that the value of $\alpha$ increases  and  approaches the critical value of $1$. Then, as
the collapse proceeds, the ensuing decreasing of the gravitational force term would eventually lead to a change of the sign
of the right hand side of (\ref{V4}). Since that would happen for small  values of the
effective inertial mass density, that would imply a strong bouncing of the sphere, even for a small absolute value of the
right hand side of (\ref{V4}).

Obviously for this effect to be of some relevance in the dynamics of the
system, evolution should proceeds in such a way that $\alpha$ approaches the critical value of $1$.
At present we may speculate that  $\alpha$ may
increase substantially (for non-negligible values of $\tau$) in a pre-supernovae event.

Indeed, at the last stages of massive star evolution, the decreasing of the opacity
of the fluid, from very high values preventing the propagation of photons
and neutrinos (trapping \cite{Ar}), to smaller values, gives rise to neutrino
radiative heat conduction. Under these conditions both $\kappa$ and $T$
could be sufficiently large as to imply a substantial increase of
$\alpha$. In fact, the values suggested in \cite{Ma} ($[\kappa] \approx
10^{37}$;
$[T] \approx 10^{13}$; $[\tau] \approx 10^{-4}$; $[\rho] \approx
10^{12}$, in c.g.s. units and Kelvin) lead to $\alpha \approx 1$. 

If it happens that  $\alpha$ gets sufficiently close to $1$ as to change the sign of the total ``force'' term in (\ref{V4}) from negative to positive, then  the
efficiency of whatever expansion mechanism of the central core at place, would be significantly enhanced. A numerical model illustrating this effect has  recently been presented
\cite{bouncing}

\end{document}